# The polarization preservation of partially coherent Hermite-Gaussian beams for multiple-degrees-of-freedom free-space communication

Ling Ji[1,2], Ai-Lin Yang[1,2], Xiao-Feng Lin[1,2], Xian-Min Jin[1,2*]

[1]State Key Laboratory of Advanced Optical Communication Systems and Networks,
Department of Physics and Astronomy, Shanghai Jiao Tong University, Shanghai 200240, China
[2]Synergetic Innovation Center of Quantum Information and Quantum Physics,
University of Science and Technology of China, Hefei, Anhui 230026, China

**Abstract**

Multiple-degrees-of-freedom free-space communication combining polarization and high-order spatial modes promises high-capacity communication channel. While high-order spatial modes have been widely exploited for dense coding and high-dimensional quantum information processing, the properties of polarization preservation of high-order spatial beams propagating in turbulent atmosphere have not been comprehensively investigated yet. Here we focus on the properties of polarization preservation of partially coherent Hermite-Gaussian beams propagating along different atmospheric turbulence paths. The analytical expressions for the polarization of partially coherent Hermite-Gaussian beams propagating through atmospheric turbulence along different paths have been derived. It is shown that the larger the coherence length is, and the larger the beam order *m, n* are, the less the polarization is changed. We find that the evolution properties of the polarization in slant-down paths through turbulent atmosphere are similar to the case in free space if the condition zenith angle $\xi<\pi/4$ is satisfied. While at a long propagation distance, evolution properties of polarization in horizontal paths of turbulent atmosphere differs much from that in free space and in slant paths. The results may allow one to choose the optimal propagation path in terms of specific applications, which is helpful for future experimental implementation of multiple-degrees-of-freedom free-space communication.

**Keywords** polarization preservation, atmospheric turbulence, laser propagation, free-space communication, quantum communication
**PACS:** 42.25.Ja, 42.68.Bz, 41.20.Jb, 42.79.Sz, 42.50.Ex

---

**Foundation items:** Supported by the National Natural Science Foundation of China (11374211), the Innovation Program of Shanghai Municipal Education Commission (14ZZ020), Shanghai Science and Technology Development Funds (No.15QA1402200) and the open fund from HPCL (No.201511-01)

*Corresponding author: E-mail: xianmin.jin@sjtu.edu.cn

**Introduction**

Photon is an ideal information carrier for long-distance communication, since photon can transmit with ultimate speed and doesn't likely couple with environment. Photonic degrees of freedom, including polarization, frequency and phase etc, have been utilized to carry information for both classical and quantum communication. Besides, high-order spatial modes of light beam, mostly known as Hermite-Gaussian (H-G) and Laguerre-Gaussian (L-G) modes which are classified according to different complete set of eigen solutions and are equivalent under cylinder lenses converter[1] are being exploited to carry more information because this degree of freedom possess very large number of dimensions[2-4]. As a typical example, orbital angular momentum derived from aforementioned high-order spatial modes has now been accepted as a new degree of photon and has shown fascinating power in high-dimensional quantum information processing[5,6].

Multiple-degrees-of-freedom free-space communication by adding other degrees of freedom into high-order spatial beams may extend the capacity of communication into an entirely new regime[7,8]. However, the properties of polarization preservation of high-order spatial beams propagating in turbulent atmosphere haven't been comprehensively investigated yet. As a step forward along this direction, here we focus on the properties of polarization preservation of partially coherent H-G beams propagating along different atmospheric turbulence paths.

The propagation of laser through atmospheric turbulence has been of considerable theoretical and practical interest for a long time, due to its important applications in optical communications and optical imaging[9]. The propagation characteristics of partially coherent electromagnetic beams through random media were studied[10-15]. The polarization of various beams propagating in turbulent atmosphere becomes a hot topic[16-21], which can be used to encode information. In 2014 Dang et al. experimentally studied the polarization fluctuations over atmospheric turbulent channels, their results are in agreement with theoretical conclusions[22].

Since Anthony E. Siegman introduced a new H-G solution, named elegant H-G

mode in 1977[23], several methods have been proposed to generate this kind of beam[24,25]. Propagation properties, such as beam intensity, mean square root width and angular spreading of H-G beams through turbulent atmosphere have been studied[26,27]. More recently, the spectral characteristics of the beam in slant turbulent atmosphere paths have attracted more and more attention[28-30].

In this letter, we have studied the effects of different turbulent paths on polarization properties of partially coherent H-G beams with various beam parameters. The results obtained in this work may deepen our understanding of the effects of different atmospheric turbulence paths on polarization characteristics of H-G beams, and provide a method of choosing the optimal propagation path efficiently with certain beam parameters, which has not been studied to our best knowledge. The physical explanations are also presented to illustrate the validity of the results.

## 1 Theory model and formulae

The field of H-G beams at the initial plane $z=0$ in the Cartesian coordinate system can be written as[27]

$$U(\boldsymbol{\rho}, z=0) = H_m\left(\frac{\sqrt{2}}{w_0}\rho_x\right) H_n\left(\frac{\sqrt{2}}{w_0}\rho_y\right) \exp\left(-\frac{\rho_x^2 + \rho_y^2}{w_0^2}\right), \qquad (1)$$

where $w_0$ denotes the waist width of Gaussian modes, $\boldsymbol{\rho} = (\rho_x, \rho_y)$ is the two dimensional position vector in the initial plane $z=0$, and $H_i(i=m, n)$ is the $i$th-order Hermite polynomial. By introducing a Gaussian term of spectral degree of coherence of $\mu(\boldsymbol{\rho}_1 - \boldsymbol{\rho}_2, 0)$, fully coherent beam can be extended into partially coherent one, which reads as

$$\mu(\boldsymbol{\rho}_1 - \boldsymbol{\rho}_2, 0) = \exp\left[-\frac{(\rho_1 - \rho_2)^2}{2\sigma_0^2}\right], \qquad (2)$$

where $\sigma_0$ is the spatial correlation length of beams in the initial plane $z=0$. Combining Eq. (1) with Eq. (2), the cross-spectral density (CSD) function of partially coherent H-G beams in the initial plane is expressed as

$$W(\boldsymbol{\rho}_1, \boldsymbol{\rho}_2, z=0) = H_m\left(\frac{\sqrt{2}}{w_0}\rho_{1x}\right) H_n\left(\frac{\sqrt{2}}{w_0}\rho_{1y}\right) \exp\left(-\frac{\rho_{1x}^2 + \rho_{1y}^2}{w_0^2}\right)$$
$$\times H_m\left(\frac{\sqrt{2}}{w_0}\rho_{2x}\right) H_n\left(\frac{\sqrt{2}}{w_0}\rho_{2y}\right) \exp\left(-\frac{\rho_{2x}^2 + \rho_{2y}^2}{w_0^2}\right) \exp\left[-\frac{(\boldsymbol{\rho}_1 - \boldsymbol{\rho}_2)^2}{2\sigma_0^2}\right], \quad (3)$$

Based on the extended Huygens-Fresnel principal[10], the CSD function of partially coherent H-G beams in the received plane $z$ is,

$$W(\boldsymbol{\rho}_1', \boldsymbol{\rho}_2', z) = \left(\frac{k}{2\pi z}\right)^2 \int_{-\infty}^{\infty}\int_{-\infty}^{\infty}\int_{-\infty}^{\infty}\int_{-\infty}^{\infty} d^2\boldsymbol{\rho}_1 d^2\boldsymbol{\rho}_2 W(\boldsymbol{\rho}_1, \boldsymbol{\rho}_2, 0)$$
$$\times \exp\left\{\frac{ik}{2z}\left[(\boldsymbol{\rho}_1' - \boldsymbol{\rho}_1)^2 - (\boldsymbol{\rho}_2' - \boldsymbol{\rho}_2)^2\right]\right\} \times \left\langle \exp\left[\psi(\boldsymbol{\rho}_1, \boldsymbol{\rho}_1') + \psi^*(\boldsymbol{\rho}_2, \boldsymbol{\rho}_2')\right]\right\rangle_m, \quad (4)$$

where $\boldsymbol{\rho}' \equiv (\rho_x', \rho_y')$ is the two dimensional position vector in plane $z$, $k$ is the wavenumber associated with wavelength by $k = \frac{2\pi}{\lambda}$, and $\langle g \rangle_m$ indicates the ensemble average of the turbulence medium statistics. Employing a quadratic approximation of Rytov's phase structure, $\langle g \rangle_m$ can be written as[31]

$$\left\langle \exp\left[\psi(\boldsymbol{\rho}_1, \boldsymbol{\rho}_1') + \psi^*(\boldsymbol{\rho}_2, \boldsymbol{\rho}_2')\right]\right\rangle_m$$
$$\approx \exp\left\{-\frac{1}{\rho_0^2}\left[(\boldsymbol{\rho}_1 - \boldsymbol{\rho}_2)^2 + (\boldsymbol{\rho}_1 - \boldsymbol{\rho}_2)(\boldsymbol{\rho}_1' - \boldsymbol{\rho}_2') + (\boldsymbol{\rho}_1' - \boldsymbol{\rho}_2')^2\right]\right\}, \quad (5)$$

where $\rho_0 = (0.545 C_n^2 k^2 z)^{-\frac{3}{5}}$ is the coherence length of spherical wave propagating in turbulence, $C_n^2$ is the structure of refractive index, which can describe the level of turbulence. So $C_n^2 = 0$ indicates the beams propagating in free space. When the beams propagate through atmospheric turbulence along the horizontal path, $C_n^2$ is constant (in this paper, we set $C_n^2 = 10^{-14} \,\mathrm{m}^{-2/3}$). Considering the slant turbulent channel (where beams propagate along a slant path, up or down), and the $\langle g \rangle_m$ can be expressed as[9]

$$\left\langle \exp\left[\psi(\boldsymbol{\rho}_1, \boldsymbol{\rho}_1') + \psi^*(\boldsymbol{\rho}_2, \boldsymbol{\rho}_2')\right]\right\rangle_m$$
$$= \exp\left\{-\frac{1}{2}\left[A_1(\boldsymbol{\rho}_1' - \boldsymbol{\rho}_2')^2 + A_2(\boldsymbol{\rho}_1' - \boldsymbol{\rho}_2')(\boldsymbol{\rho}_1 - \boldsymbol{\rho}_2) + A_3(\boldsymbol{\rho}_1 - \boldsymbol{\rho}_2)^2\right]\right\}, \quad (6)$$

with

$$A_1 = 3.2796 k^2 l_0^{-\frac{1}{3}} \sec(\xi) \times \int_{h_0}^{H} C_n^2(h)(1-\eta)^2 dh, \tag{7a}$$

$$A_2 = 3.2796 k^2 l_0^{-\frac{2}{3}} \sec(\xi) \times \int_{h_0}^{H} 2 C_n^2(h)\eta(1-\eta) dh, \tag{7b}$$

$$A_3 = 3.2796 k^2 l_0^{-\frac{1}{3}} \sec(\xi) \times \int_{h_0}^{H} C_n^2(h)\eta^2 dh, \tag{7c}$$

where $\xi$ denotes the zenith angle, $z$ is the propagation distance, $h$ is the altitude from the ground ($h = z\cos(\xi)$), and $H$ indicates the altitude between the initial plane and received plane. $\eta = 1 - h/H$ and $\eta = h/H$ correspond to the beams propagating through atmospheric turbulence along the slant-up path and the slant-down path, respectively. $C_n^2(h)$ is an altitude-dependent structure index constant[9]

$$\begin{aligned}C_n^2(h) &= 0.00594 \left(\frac{v}{27}\right)^2 \times (10^{-5} h)^{10} \times \exp\left(-\frac{h}{1000}\right) \\ &+ 2.7 \times 10^{-16} \exp\left(-\frac{h}{1500}\right) + C_n^2(0)\exp(-h/100)\end{aligned}, \tag{8}$$

where $v$ denotes the root-mean-square wind speed (in this paper, we set $v = 2 \cdot 1 \text{m/s}$). To obtain the analytical result, two variables of integration are introduced as $\boldsymbol{u} = \frac{\boldsymbol{\rho}_1 + \boldsymbol{\rho}_2}{2}$, $\boldsymbol{v} = \boldsymbol{\rho}_2 - \boldsymbol{\rho}_1$. Assuming $\boldsymbol{\rho}'_1 = \boldsymbol{\rho}'_2 = \boldsymbol{\rho}'$ and substituting Eq. (6), Eq. (7) and Eq. (8) into Eq. (4), we obtain the CSD function of partially coherent H-G beams in the received plane $z$.

$$\begin{aligned}W(\boldsymbol{\rho}', \boldsymbol{\rho}', z) &= I(\boldsymbol{\rho}', z) \\ &= \left(\frac{k}{2\pi z}\right)^2 \iint d^2 u \iint d^2 v \exp\left(-\frac{2u^2}{w_0^2}\right)\exp\left(-\frac{v^2}{\varepsilon^2}\right)\exp\left(-\frac{ik}{z}\boldsymbol{u}\boldsymbol{v}\right)\exp\left(\frac{ik}{z}\boldsymbol{\rho}'\cdot\boldsymbol{v}\right) \\ &\times H_m\left[\frac{\sqrt{2}}{w_0}\left(u_x - \frac{v_x}{2}\right)\right] H_m\left[\frac{\sqrt{2}}{w_0}\left(u_x + \frac{v_x}{2}\right)\right] H_n\left[\frac{\sqrt{2}}{w_0}\left(u_y - \frac{v_y}{2}\right)\right] H_n\left[\frac{\sqrt{2}}{w_0}\left(u_y + \frac{v_y}{2}\right)\right]\end{aligned}, \tag{9}$$

where $\frac{1}{\varepsilon^2} = \frac{1}{2w_0^2} + \frac{1}{2\sigma_0^2} + \frac{1}{\rho^2}$, $\frac{1}{\rho^2}$ denotes the phase change caused by turbulence, $\rho = \rho_0$ and $\rho = \sqrt{\frac{2}{A_3}}$ represent horizontal path and slant path, respectively.

With the help of the generalized Laguerre polynomial and the integral formula

$$L_m^{\alpha+\beta+1}(x+y) = \sum_n^m L_n^{\alpha}(x) L_{m-n}^{\beta}(y), \tag{10}$$

$$L_n^0(x) = \sum_{m=0}^n (-1)^m \binom{n}{n-m} \frac{x^m}{m!}, \tag{11}$$

$$\int_{-\infty}^{\infty} x^n \exp(-px^2 + 2qx) dx = n! \exp\left(\frac{q^2}{p}\right) \sqrt{\frac{\pi}{p}} \left(\frac{q}{p}\right)^n \sum_{k=0}^{E\left[\frac{n}{2}\right]} \frac{1}{(n-2k)!k!} \left(\frac{p}{4q^2}\right)^k, \tag{12}$$

We can obtain

$$W(\boldsymbol{\rho}',\boldsymbol{\rho}',z) = I(\boldsymbol{\rho}',z) = \left(\frac{k}{4z}\right)^2 \frac{w_0^2}{2a} S_x S_y, \tag{13}$$

with

$$a = \frac{k^2 w_0^2}{8z^2} + \frac{1}{\varepsilon^2}, \quad b_x = \frac{ik\rho_x'}{2z}, \quad b_y = \frac{ik\rho_y'}{2z}, \quad d = \frac{k^2 w_0^2}{4z^2} + \frac{1}{w_0^2},$$

$$S_x = 2^m m! \exp\left(\frac{b_x^2}{a}\right) \sum_{l=0}^m (-1)^l (2l)! \binom{m}{m-l} \frac{d^l}{l!} \left(\frac{b_x}{a}\right)^{2l} \sum_{k=0}^l \frac{1}{(2l-2h)!k!} \left(\frac{a}{4b_x^2}\right)^k,$$

$$S_y = 2^n n! \exp\left(\frac{b_y^2}{a}\right) \sum_{l=0}^m (-1)^l (2l)! \binom{n}{n-l} \frac{d^l}{l!} \left(\frac{b_y}{a}\right)^{2l} \sum_{k=0}^l \frac{1}{(2l-2h)!k!} \left(\frac{a}{4b_y^2}\right)^k, \tag{14}$$

$I(\boldsymbol{\rho}',z)$ in Eq. (13) denotes the intensity of partially coherent H-G beams

In order to study the effects of different paths in turbulence (i.e., slant-up, slant-down, horizontal) with beam parameters (i.e., beam order, spatial correlation length, zenith angle of path) on the degree of polarization of partially coherent H-G beams, and having compared with the effects in free space, we introduce the polarization matrix of partially coherent H-G beams[11], which reads as

$$\begin{bmatrix} W_{xx}(\boldsymbol{\rho}_1',\boldsymbol{\rho}_2') & W_{xy}(\boldsymbol{\rho}_1',\boldsymbol{\rho}_2') \\ W_{yx}(\boldsymbol{\rho}_1',\boldsymbol{\rho}_2') & W_{yy}(\boldsymbol{\rho}_1',\boldsymbol{\rho}_2') \end{bmatrix}, \tag{15}$$

with

$$W_{xx}(\boldsymbol{\rho}_1',\boldsymbol{\rho}_2') = \gamma_{xx} U_x(\boldsymbol{\rho}_1') U_x(\boldsymbol{\rho}_2'), \tag{16a}$$

$$W_{yy}(\boldsymbol{\rho}_1',\boldsymbol{\rho}_2') = \gamma_{yy} U_y(\boldsymbol{\rho}_1') U_y(\boldsymbol{\rho}_2'), \tag{16b}$$

$$W_{xy}(\boldsymbol{\rho}_1',\boldsymbol{\rho}_2') = \gamma_{xy} U_x^*(\boldsymbol{\rho}_1') U_y(\boldsymbol{\rho}_2') = W_{yx}^*(\boldsymbol{\rho}_1',\boldsymbol{\rho}_2'), \tag{16c}$$

where $\gamma_{ij}$ ($i,j=x,y$) is the normalized cross correlation between the two components at the same point[32], and their absolute values satisfy $0 \leq |\gamma_{ij}| \leq 1$. The value 0 and 1 correspond to the totally uncorrelated and completely correlated, respectively. The degree of polarization $P(\boldsymbol{\rho}',z)$ of partially coherent H-G beams at received plane $z$ is given as[11]

$$P(\boldsymbol{\rho}',z) = \sqrt{1 - \frac{4 det\left[\hat{W}(\boldsymbol{\rho}',\boldsymbol{\rho}',z)\right]}{\left\{trace\left[\hat{W}(\boldsymbol{\rho}',\boldsymbol{\rho}',z)\right]\right\}^2}}, \quad (17)$$

where trace[$W$] represents the trace of CSD matrix. $W_{ij}(\boldsymbol{\rho}_1',\boldsymbol{\rho}_2')$ can be derived from $W(\boldsymbol{\rho}_1',\boldsymbol{\rho}_2')$[33]. Its expression is

$$W_{ij}(\boldsymbol{\rho}',\boldsymbol{\rho}',z) = \gamma_{ij}\left(\frac{k}{4z}\right)^2 \frac{w_0^2}{2a} S_{xij} S_{yij}, \quad (18)$$

$$S_{xij} = 2^m m! \exp\left(\frac{b_x^2}{a_{ij}}\right) \sum_{l=0}^{m}(-1)^l (2l)! \binom{m}{m-l}\frac{d^l}{l!}\left(\frac{b_x}{a_{ij}}\right)^{2l} \sum_{k=0}^{l}\frac{1}{(2l-2h)!k!}\left(\frac{a_{ij}}{4b_x^2}\right)^k, \quad (19a)$$

$$S_{yij} = 2^n n! \exp\left(\frac{b_y^2}{a_{ij}}\right) \sum_{l=0}^{n}(-1)^l (2l)! \binom{n}{n-l}\frac{d^l}{l!}\left(\frac{b_y}{a_{ij}}\right)^{2l} \sum_{k=0}^{l}\frac{1}{(2l-2h)!k!}\left(\frac{a_{ij}}{4b_y^2}\right)^k, \quad (19b)$$

$$a_{ij} = \frac{k^2 w_0^2}{8z^2} + \frac{1}{\varepsilon_{ij}^2}, \quad \frac{1}{\delta_{ij}^2} = \frac{1}{\rho_0^2} + \frac{1}{\sigma_{0ij}^2}, \quad b_x = \frac{ik\rho_x'}{2z}, \quad b_y = \frac{ik\rho_y'}{2z}, \quad d = \frac{k^2 w_0^2}{4z^2} + \frac{1}{w_0^2}. \quad (19c)$$

where $i=x,y; j=x,y$. Substituting Eq. (18) into Eq. (17), the expression of polarization degree of partially coherent H-G beams can be obtained.

## 2  Numerical examples and analysis

Numerical examples are given by using Eq. (17) to show the effects of different propagation paths in atmospheric turbulence and in free space on the polarization of partially coherent H-G beams, as depicted in Fig. 1-5. The evolution behaviors of average intensity distribution are characterized by Eq. (13). Figure 1 shows the behaviors of the normalized average intensity $I(\rho',z)/I_{max}(\rho',z)$ and polarization distribution versus the

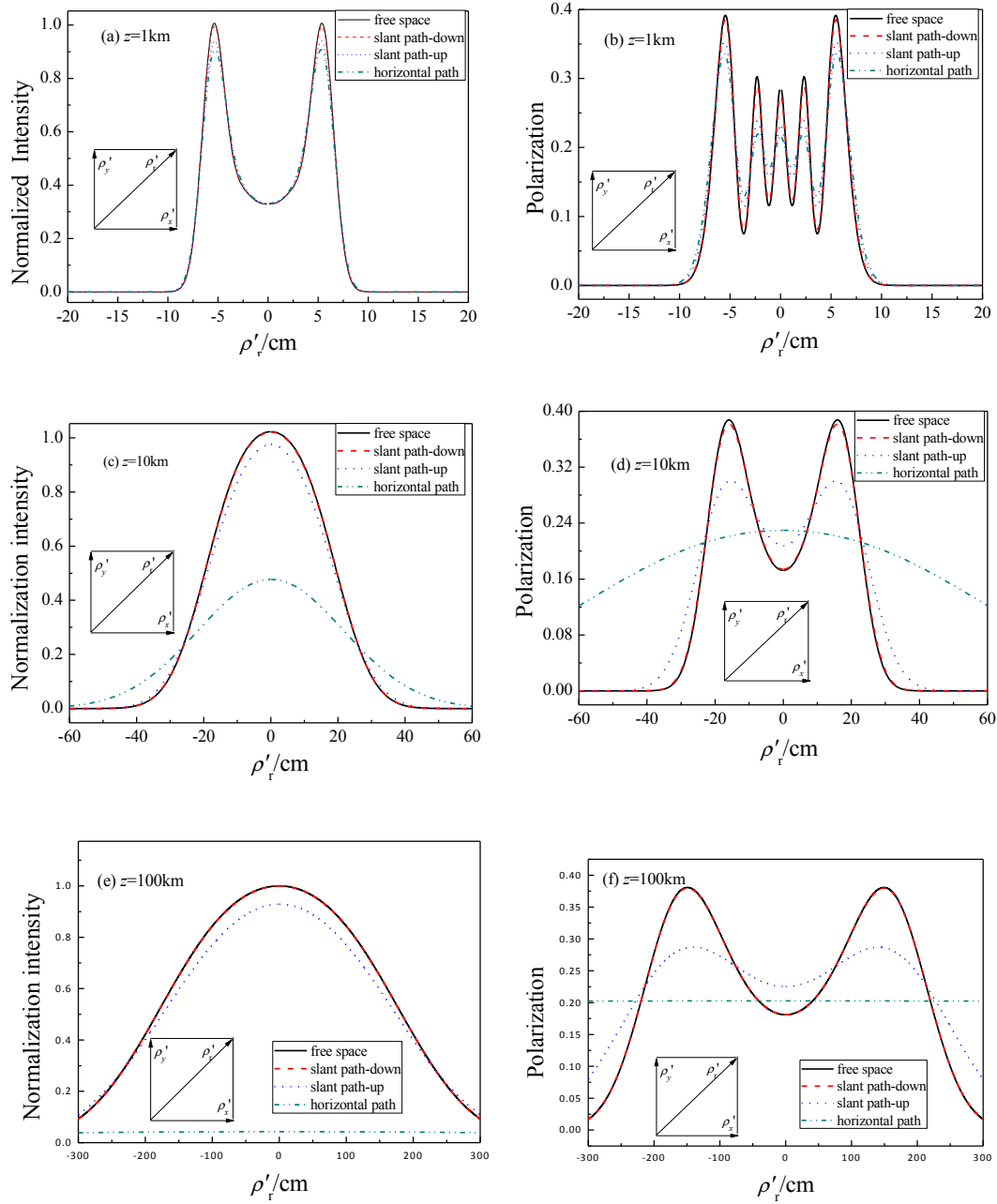

Fig. 1 Intensity and polarization of a partially coherent H-G beam at several propagation distances for different propagation paths in atmospheric turbulence: (a) (c) (e) intensity distribution (b) (d) (f) polarization distribution.

slanted axis at some specific propagation distances, where $I_{max}(\rho',z)$ denotes the maximum average intensity at the received plane $z$. The calculation parameters are $\lambda = 800$ nm, $\sigma_{0xx} = \sigma_{0yy} = 1$ cm, $\sigma_{0xy} = 2$ cm, $\gamma_{xx} = \gamma_{yy} = 0.5$, $\gamma_{xy} = 0.1$, $w_0 = 3$ cm, $m = n = 4$, $\xi = \dfrac{\pi}{3}$, $C_n^0(0) = 10^{-14}$ m$^{-2/3}$. From Fig.1a, at a short propagation distance,

$z$=1km for example, the average intensity profile changes to Gaussian profile with a dip (the initial intensity distribution is Hermite-Gaussian profile, which is not shown), and finally evolves into Gaussian profile when the propagation distance is sufficiently long. The polarization distributions of partially coherent H-G beams propagating through different atmospheric turbulence paths are similar but not the same as intensity behaviors, which can be seen from Fig. 1b, d, f. The polarization distributions remain Gaussian profile with a dip for a beam propagating through turbulence in slant paths at a long distance. The effect of turbulence on beam properties is negligible in short range propagation, and the slant-down path in turbulence can be substituted for free space at any distance.

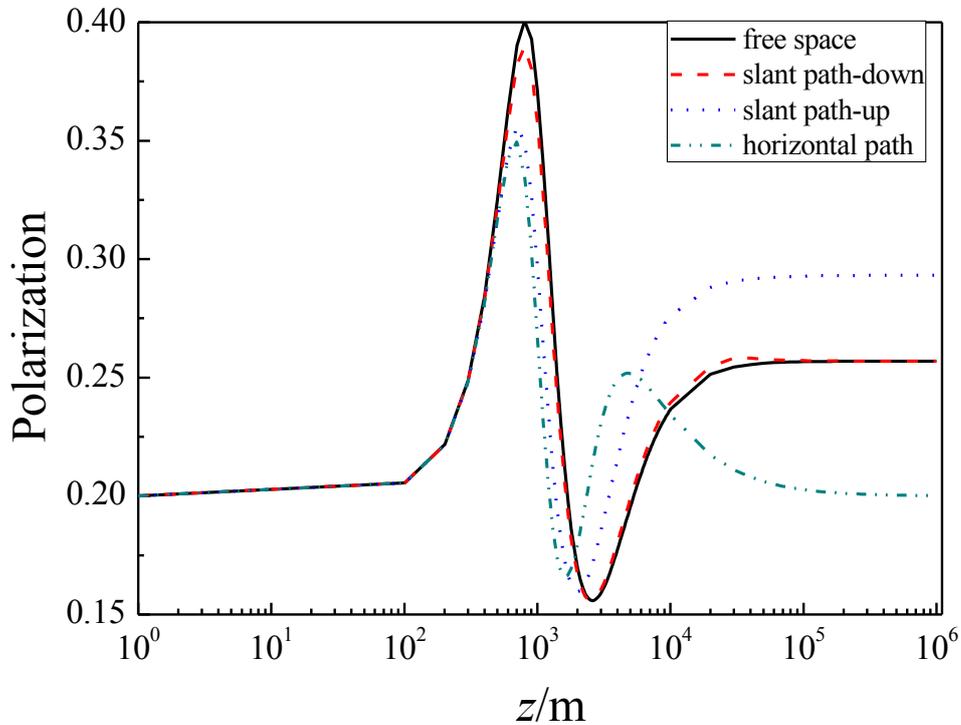

Fig. 2 Polarization of a partially coherent H-G beam versus the propagation distance for different turbulence paths.

Polarization properties $P(0,0,z)$ of a partially coherent H-G beam propagating through atmospheric turbulence along different paths are represented in Fig. 2, and $m=n=2$, the other calculation parameters are the same as in Fig. 1. When the propagation distance is short, e.g., $z<100$m, the polarization remains almost unchanged. The degrees of polarization tend to some particular values after propagating enough long distance and retain the values as the beams propagate further. The degree of polarization returns to its initial value after it propagates through turbulence along horizontal path over a sufficiently long distance. This conclusion proves the feasibility of long-range distance laser communication based on beam polarization.

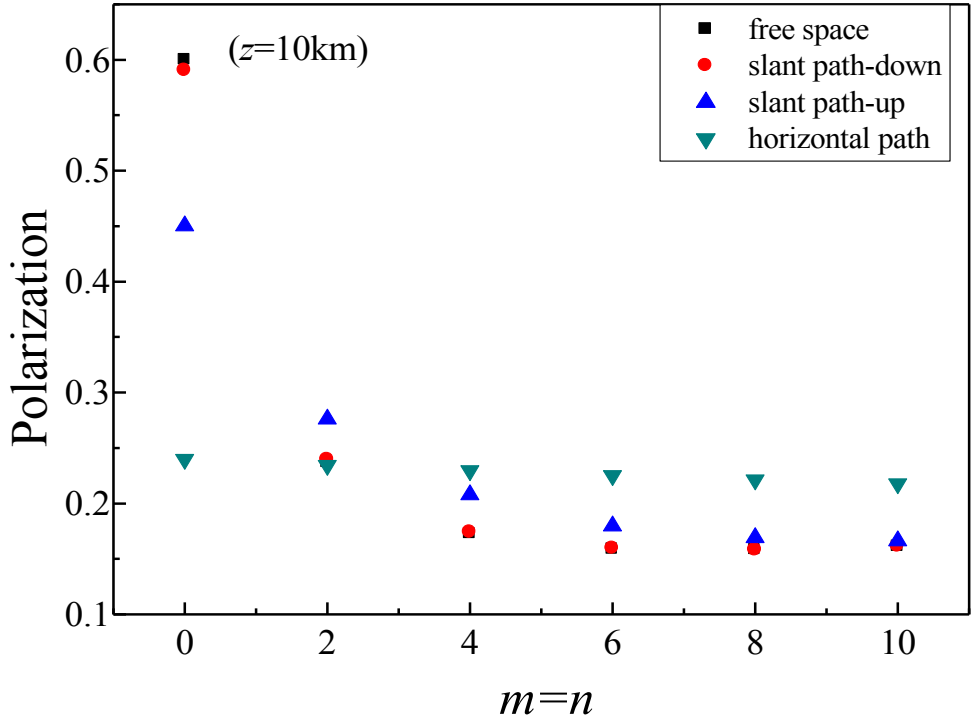

Fig. 3 Polarization of a partially coherent H-G beam versus the beam orders $m$, $n$ for different turbulence paths.

The polarization of partially H-G beams at the plane $z=10$ km versus the beam orders is plotted in Fig. 3, the other calculation parameters are the same as those in Fig. 2. It is shown that the polarization of partially H-G beams is dependent on the propagating path in turbulence. The differences of polarization between different propagation paths in turbulence decrease as the beam orders $m$, $n$ increase, e.g., $m=n=0$, $P_{free\ space}$, $P_{slant\ path-down}$, $P_{slant\ path-up}$, $P_{horizontal\ path}$ are 0.600, 0.590, 0.450 and 0.239, while $m=n=10$, $P_{free\ space}$, $P_{slant\ path-down}$, $P_{slant\ path-up}$, are 0.161, 0.161, 0.166 and 0.218, respectively. Therefore, the polarization of partially H-G beams with larger $m$, $n$ is less affected by atmospheric turbulence than that of partially H-G beams with smaller $m$, $n$. Beams with high orders $m$,

$n$ propagating through slant paths can be simplified in that case in free space.

The polarization of partially H-G beams with various propagation paths in turbulence for different values of coherence length is plotted in Fig. 4, the other calculation parameters are the same as those in Fig. 2. From Fig. 4 it is seen that the polarization changes dramatically when the coherence length is small, e.g., it rapidly increases to 0.7 for the beam of $\sigma_{0xx}=\sigma_{0yy}=1$ mm, $\sigma_{0xy}=2$ mm. The polarization of the beam with larger coherence length varies slightly with the propagation distance. Therefore, the polarization of partially coherent H-G beams with a larger coherence length is less affected by atmospheric turbulence than that of partially coherent H-G beams with a smaller coherence length. The slant-up path can be simplified to the horizontal path at a long propagation distance if the beam possesses a large coherence length.

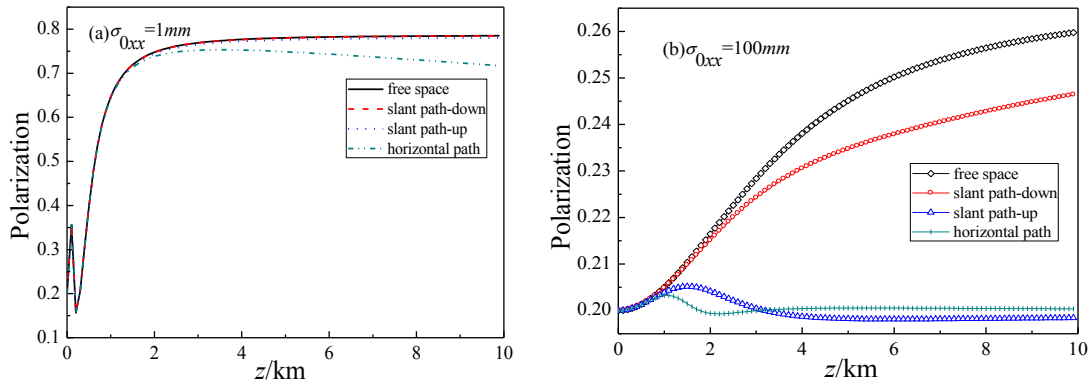

Fig. 4 Polarization of a partially coherent H-G beam versus the propagation distance for different turbulence paths. (a) $\sigma_{0xx}=1\,\text{mm}$ (b) $\sigma_{0xx}=100\,\text{mm}$

Figure 5 gives the polarization of partially coherent H-G beams at several planes versus zenith angle $\xi$. The phenomena of polarization changes versus $\xi$ in slant propagation paths are interesting. When the propagation distance is short, the polarization decreases as the $\xi$ increases. The polarization varies non-monotonously with zenith angle at a long distance, e g., at $z$=20km, the values of polarization reaches maximum 0.291 and 0.288 for $\xi$=72 and 88 degree, corresponding to the slant-up and slant-down paths, respectively. The polarization decreases dramatically when the $\xi$ is close to $\frac{\pi}{2}$, as a slight change in $\xi$ causes a huge alternation of structure of refractive index in turbulence along slant paths, especially along a slant-down path. The results can be physically explained as follows. The relation between the structure of refractive index and altitude is characterized by Eq. (8), some particular altitudes are shown in Table 1. From the table, we can find that the turbulence plays a dominant role in low altitudes. Therefore, when the propagation distance is sufficient long, if $\xi$ is closer to $\frac{\pi}{2}$, a slight increase in $\xi$ causes a noticeable ratio of low altitude path to high altitude path, while if the zenith angle is small, the ratio is not so sensitive to the zenith angle.

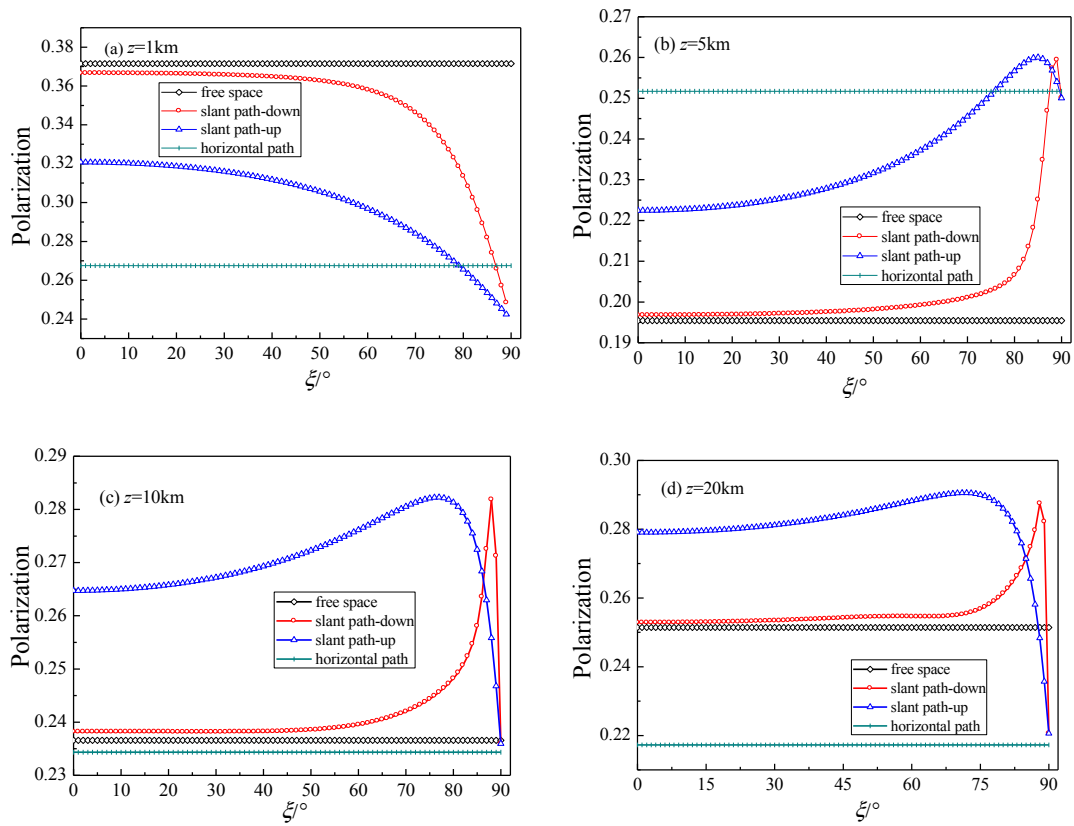

Fig. 5 Polarization of a partially coherent H-G beam versus zenith angle at several propagation distances for different turbulence paths.

Table 1 The relation between altitude and refractive index structure

| Altitude(m) | structure of refractive index ($m^{-2/3}$) |
|---|---|
| 0 | $10^{-14}$ |
| 100 | $3.93\times10^{-15}$ |
| 200 | $1.59\times10^{-15}$ |
| 256 | $10^{-15}$ |
| 300 | $7.19\times10^{-16}$ |
| 800 | $1.62\times10^{-16}$ |
| 1485 | $10^{-16}$ |

## 3  Conclusion

In this paper, the polarization properties of partially coherent H-G beams propagating through atmospheric turbulence along different paths have been studied both analytically and numerically. It has been found that the intensity and polarization distribution of the partially coherent H-G beams in atmospheric turbulence undergo several stages of evolution, which depend on the turbulence paths and beam parameters. The larger the spatial correlation lengths and the larger the beam orders *m*, *n* are, the less the beam polarization is changed provided that the propagation distance is sufficiently long. Partially coherent H-G beams propagating through the turbulence along a slant-down path can be simplified in free space if the condition $\xi\in[0,\frac{\pi}{4}]$ is satisfied. The beams propagating through the turbulence along a slant-up path can be substituted for a horizontal path if the correlation length is large enough. The polarization is dramatically dependent on the zenith angle when the zenith angle is close to $\frac{\pi}{2}$. The physical interpretation has been given to show the validity of our results. Our results allow one to choose the optimal propagation path in terms of specific applications, and would be useful for future experimental implementation of multiple-degrees-of-freedom free-space communication.